# Efficient Reconstruction of Free Breathing Under-Sampled Cardiac Cine MRI

Abdul Haseeb[1] Ahmed, Ijaz M. Qureshi[1], Jawad Ali Shah[2], and Hammad Omer[3].

1. haseeb@mail.au.edu.pk. imqureshi@mail.au.edu.pk. Department of Electrical Engineering, Air University Islamabad, Pakistan.
2. Jawad.shah@iiu.edu.pk. Department of Electronic Engineering Faculty of Engineering & Technology International Islamic University, Islamabad, Pakistan.
3. Hammad.omer@comsats.edu.pk. Department of Electrical Engineering, COMSATS Institute of information technology, Islamabad, Pakistan.

Correspondence to: Abdul Haseeb Ahmed; email: haseeb@mail.au.edu.pk.

***Abstract*—Respiratory motion can cause strong blurring artifacts in the reconstructed image during MR acquisition. These artifacts become more prominent when use in the presence of under-sampled data. Recently, compressed sensing (CS) is developed as an MR reconstruction technique, to recover good quality images from the compressive k-space samples. To maximize the benefits of CS in free breathing data, it is understandable to use CS with the motion corrected images. In this paper, we have developed a new CS based motion corrected image reconstruction technique. In this two-stage technique, we use similarity measure to sort the motion corrupted data into different respiratory states. Then, we use a new reconstruction algorithm, which iteratively performs reconstruction and motion correction. The performance of the proposed method is qualitatively and quantitively evaluated using simulated data and clinical data. Results depict that this method performs the better reconstruction of respiratory motion corrected cardiac cine images as compared to the CS based reconstruction method.**

## I. Introduction

The development of novel imaging techniques, such as magnetic resonance imaging (MRI), helps to assess and monitor the cardiovascular diseases (CVD) so that to provide the effective treatment. It is necessary to acquire accelerated MR images by using fast acquisition sequences and/or by reducing the number of data acquisitions [1] [2]. As the former technique is inherently limited by many constraints, much research efforts are put to the latter approach. To accelerate the data acquisition, under-sampling method is used, which results in the noisy images. Efficient reconstruction techniques are required to remove the noise in the images. To reconstruct fully sampled images from under-sampled data, parallel imaging [3-5], and compressed sensing [6-9], are used. Multiple receiver coils are employed in parallel imaging for data acquisition, where the data from each coil is weighted with the sensitivity profile of the coil. The redundancy of the multi-coil data is exploited in the reconstruction process to obtain fully sampled images. A drawback of parallel imaging is that, as the acceleration rate increases the reconstruction artifacts increase rapidly. Compressed sensing (CS) has been used as a technique that exploits sparsity in the acquired data to reconstruct the MR data from highly under-sampled measurement. Compressed Sensing (CS) is successfully used in the cardiac cine MRI, where data is compressible in the temporal dimension [8].

Conventionally, breath-held setting is used to avoid respiratory motion artifacts in the cardiac cine MRI. However, it is not comfortable for cardiac patients to hold their breath for data acquisition. Researchers are now doing efforts to obtain motion free cardiac cine MRI [10, 11]. Motion artifacts such as blurring and ghosting are prominent in the cardiac cine MRI due to patient movements during data acquisition and breathing motion.

Problems related to the reconstruction of free-breathing cardiac cine MRI are:

- K-space samples are obtained at distinctive respiratory motion states or breathing positions. These samples, when use in reconstruction, can cause motion artifacts in the recovered images.
- Due to motion artifacts, the sparsity level of the images in the sparse domain reduces. Thus, adding a constraint to achieve the higher acceleration rates with CS reconstruction.

Recently, some methods have been proposed, which use CS reconstruction with motion correction techniques. Otazo et al. proposed a method which combines parallel imaging and CS with 1D translational respiratory motion correction [12].

Motion corrected CS for dynamic cardiac MRI has been proposed by Usman et al.[13]. This method utilizes a generalized motion correction framework which is used to recover the motion corrected images[14]. This framework modeled the transformation to the acquired motion corrupted k-space data at different states of motion from the motion free image. Feng et al. suggested the technique, which combined the parallel imaging, compressed sensing, and golden-angle radial sampling to acquire high spatial-temporal resolution with motion robustness [15]. In another work, Feng et al. utilized the extra motion-state dimensions in the under-sampled radial k-space data to recover the MR images [16]. It was shown that this method separated the respiratory and cardiac motions from cardiac cine MRI and reduced blurring.

In this paper, we propose a two-stage approach, which incorporate a motion corrected CS imaging framework for free-breathing cardiac cine MRI, which utilizes a motion correction formulation directly in the CS based imaging reconstruction. First, free breathing k-space data are binned to the different respiratory states by using a novel similarity based method.

Then, a propose reconstruction algorithm is used, to estimate the missing samples and the displacement matrix iteratively, which corrects motion artifacts and reconstructs cine images from under-sampled free breathing k-space data. The number of acquired variable density k-space samples in each motion state is below the Nyquist rate. The proposed motion correction reconstruction method outperforms the conventional CS based reconstruction method.

## II. EXPLAINATION

The effect of motion artifacts during the acquisition of k-space data can be illustrated using following formulation. Let $x_0$ be the motion-free image to be reconstructed, and $y_m$ be the motion corrupted k-space data acquired at $m$ respiratory state. The motion corrupted k-space data $y_m$ is defined as:

$$y_m = A_m F U_m x_0 \quad (1)$$

where, $U_m$ is a displacement matrix that warps the pixels in reference respiratory state to the $m_{th}$ respiratory state, $F$ is the 2D Fourier encoding matrix that transforms the image to the k-space data, and $A_m$ is the k- space under-sampling operator at $m$ respiratory state. To apply motion corrected rapid imaging formulation for general motion correction, the k-space profiles in the acquired free-breathing cardiac MR data have to be associated to different motion states such that respiratory motion within each motion state can be assumed small enough not to cause motion artifacts in the images reconstructed from the acquired data. The motion between different motion states can be estimated by first reconstructing the images at each motion state and then registering these images to get the motion information $\boldsymbol{U}$. The most common respiratory motion state is chosen as the reference, and all the reconstructed images are non-rigidly registered to this reference. The registrations provide an estimate of motion and the motion corrected CS imaging can be used to reconstruct images free of respiratory motion.

There are two stages involved in the proposed scheme. In the first stage, free breathing k-space is transformed into under sampled images. A variable density under sampling scheme is used to acquire the majority of samples around the center of the k-space. Therefore, we measure the similarity of the images using mutual information based similarity function. The k-space data from the same respiratory states are binned with the help of this similarity measure. The k-space data from the different respiratory states are reconstructed using compressed sensing based recovery algorithm. In second stage, we iteratively improve the displacement matrix $\boldsymbol{U}$ and the recover the missing samples in the acquired data. The reconstructed images from the stage one is registered to the reference state to obtain the displacement matrix $\boldsymbol{U}$. Then the displacement matrix $\boldsymbol{U}$ is used in the CS based motion correction framework as mentioned in equation 1, to recover the missing samples. Finally, the motions free images $x_0$ are obtained. The displacement information between respiratory states can be obtained by first reconstructing the images at each state and then applying image registration method, to get the motion information $\boldsymbol{U}$. The most common respiratory state is chosen as the reference, and all the CS reconstructions are non-rigidly registered to the reference state. Each stage is explained below in detail and is shown in Figure 1.

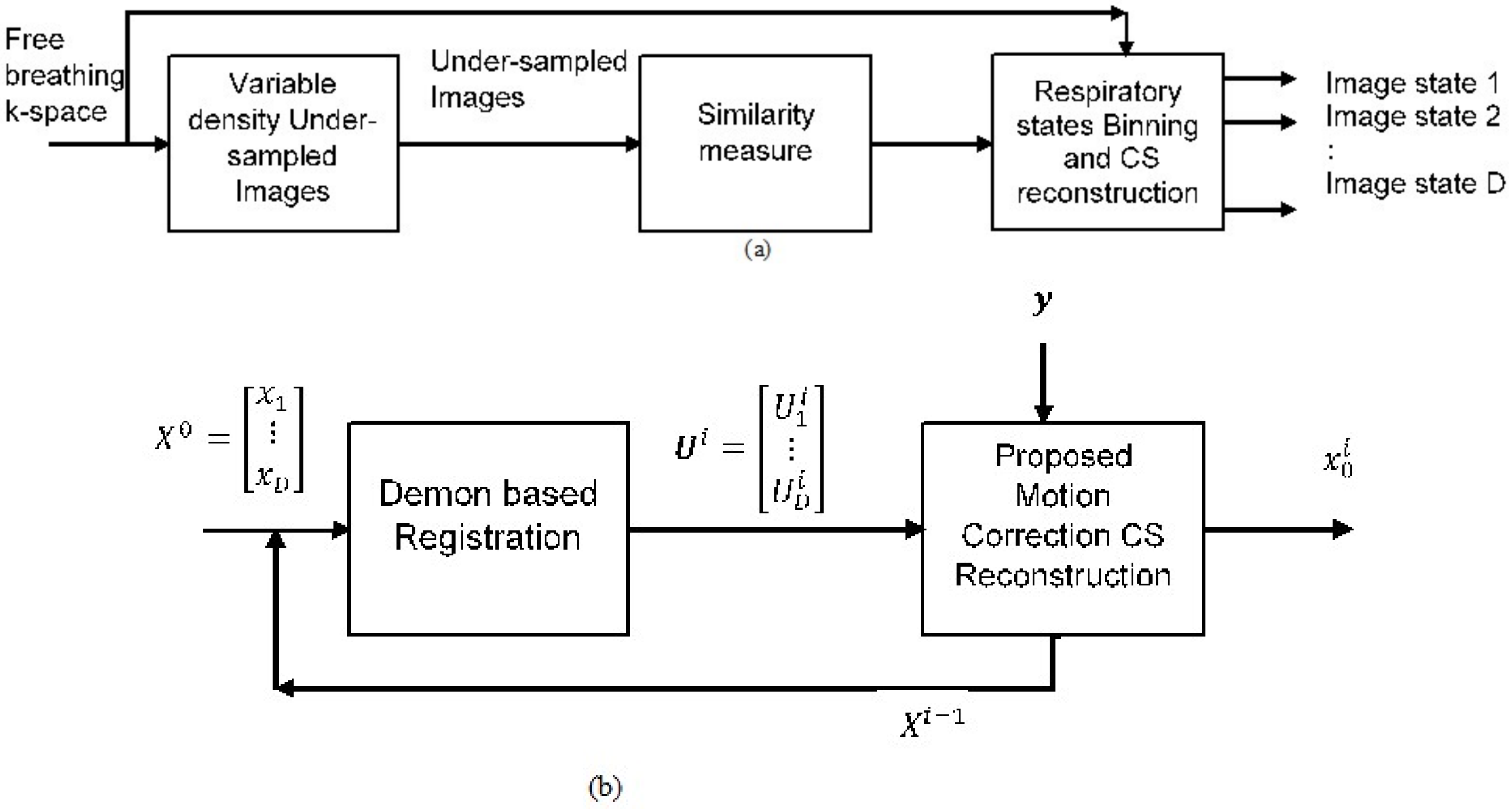

*Figure 1: Proposed block diagram of motion correction based cardiac cine reconstruction*

## A. First Stage

### 1. Variable density under sampled images

The acquisition of the free breathing under sampled k-space data is done using variable density technique, in which more samples are acquired around the center of k-space and less samples are obtained in the periphery. Fourier transform is then used to obtain the under sampled images from the free breathing under sampled k-space data.

### 2. Similarity Measure

SSF based reconstructed images, which belong to the same respiratory states, are combined using a similarity function. Similarity function is commonly used as a cost function in image alignment. It measures the amount of similarity between two images. In rigid transformation, as in the case of breathing artifacts in the images, this function provides the amount translational factor between two images. The two images will be more similar, if there is small translation or rotation between them. In this work, information theory based mutual information between two images is used as a similarity measure. Mutual information (MI) is defined by

$$MI = \sum_{x,y} p(x,y) \log \frac{p(x,y)}{p(x)p(y)} , \qquad (2)$$

Where, $p(x,y)$ is the joint probability distribution, $p(x)$ and $p(y)$ are the probability distributions of the individual images. Mutual information does not assume a linear relationship between the pixel values of the two images, but instead assumes that the co-occurrence of the most probable values in the two images is maximized at registration. The respiratory signal is obtained using similarity based method as shown in Figure 2.

### 3. Proposed Reconstruction for dynamic MRI

Free-breathing cardiac data for a number of cardiac cycles is acquired using variable density Cartesian sampling. Acquired data belongs to a respiratory state, within each R-R interval, and is retrospectively given to N cardiac phases based on an external electrocardiogram (ECG) signal. For each respiratory state, the k-space data lies in $k_x$–$k_y$–$n$ space, where $n$ refers to the cardiac phase. In the first stage, compressed sensing based image reconstruction is employed on k-space data acquired at each cardiac cycle [8]. As cardiac cine MR images are sparse in the temporal domain, so images are reconstructed by minimizing the following cost function:

$$f(x) = \frac{1}{2}\left\| y - A\ Fx \right\|_2^2 + \beta \left\| Wx \right\|_1 , \qquad (3)$$

Where y is the acquired k-space data, x is the desired image, W is the sparsifying transform in the temporal dimension and β is a regularization parameter that controls the tradeoff between sparsity and data consistency. Equation 3 is minimized using separable surrogate functional based reconstruction method proposed in [17]. The SSF based algorithm uses the following equation at each iteration:

$$\hat{x}_{i+1} = W^H \left\{ S\left( W\left( \frac{1}{c} F^H \left( y - Fx_i \right) + x_i \right), \frac{\beta}{c} \right) \right\}, \qquad (4)$$

Where S(.) is the soft thresholding function:

$$S(x, \frac{\beta}{c}) = \begin{cases} \frac{x}{|x|}(|x| - \frac{\beta}{c}), & |x| > \frac{\beta}{c} \\ 0, & otherwise \end{cases}$$

In first stage, CS based separable surrogate function (SSF) reconstruction is performed for each respiratory state independently, with the *x–y–f* space as the sparse representation. This step reconstructs N cardiac phases for each respiratory motion state.

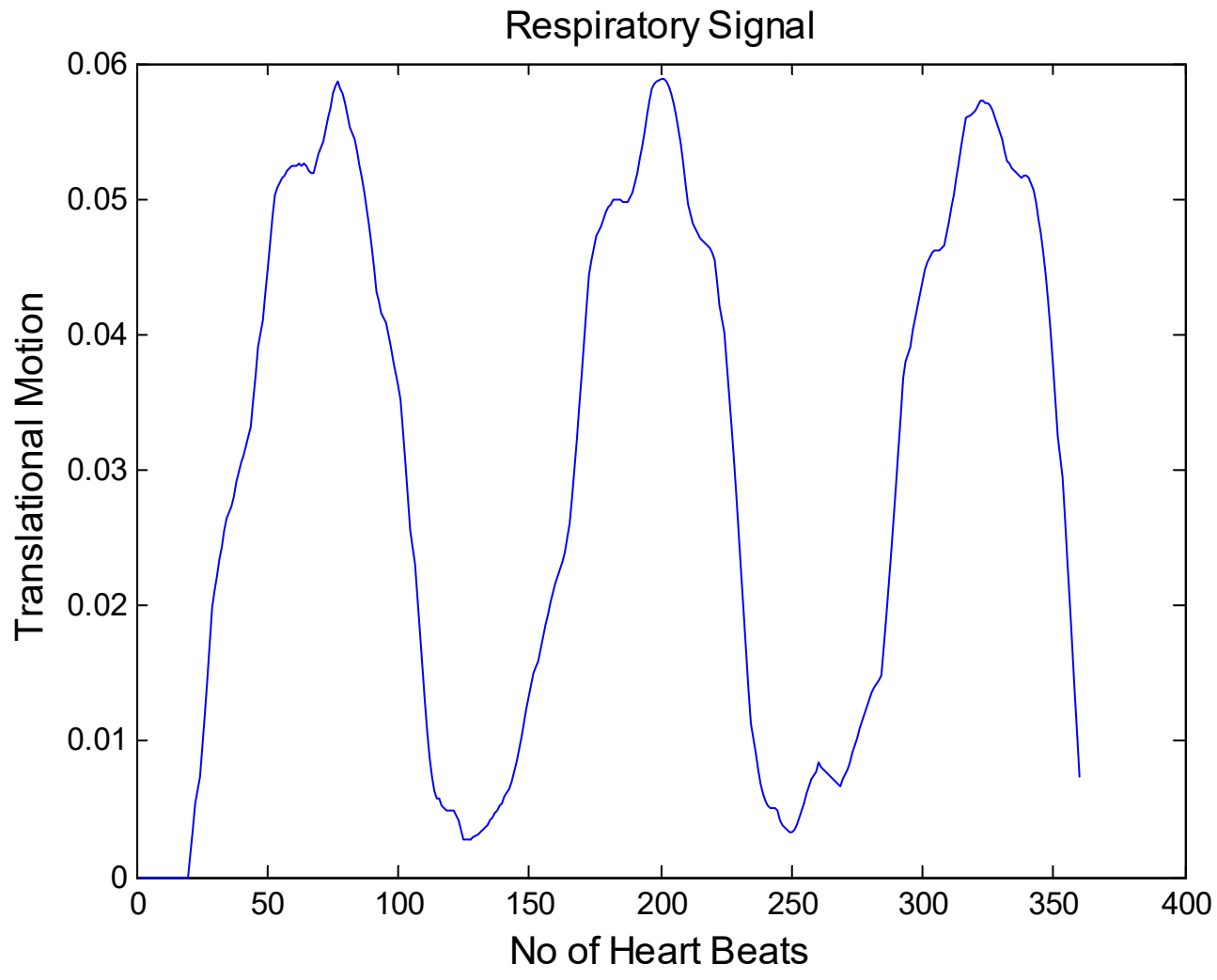

*Figure 2: Respiratory signal obtained using similarity measure based method*

## B. Second Stage

### 1) Demon based registration

Compressed sensing reconstructed cardiac cine images at the reference respiratory state is registered to the other respiratory states using the demon non-rigid registration algorithm [18] [19]. It is a high precision based registration which exploits pixel velocities caused by edge based forces. The resultant pixel velocity is filtered by Gaussian kernel to achieve global registration. Pixel-wise displacement fields for every cardiac phase within each respiratory state are obtained from the registration process. These pixel based displacement fields are used to construct the two dimensional motion matrix $U_m$ at a specific respiratory state $m$. Demon based registration method is given as:

$$U_m = \frac{(x_m - x_1)\nabla x_1}{|\nabla x_1|^2 + \alpha (x_m - x_1)^2} + \frac{(x_m - x_1)\nabla x_m}{|\nabla x_m|^2 + \alpha (x_m - x_1)^2} \qquad (5)$$

### 2) Proposed reconstruction of free breathing dynamic MRI

Using the estimated motion matrix (from registration process) $U$ and the acquired data y from all the respiratory states, the final motion corrected images are reconstructed using the following equation:

$$f(x) = \frac{1}{2}\left\| y - A_m F\ U_m x \right\|_2^2 + \beta \left\| Wx \right\|_1 \qquad (6)$$

Where x is the motion corrected image. Equation 5 is minimized using separable surrogate functional based algorithm. The algorithm uses the following steps shown in Table 1.

This approach enforces true data consistency, and updates only the missing k-space points. Each respiratory state corrupted data gives motion free images after this stage. Final cardiac cine MR images are obtained by using sum of square of the images obtained from the third stage.

Table 1 Proposed Algorithm

**Input:** Under sampled k-space data vector $y \in \mathbb{C}^M$ where $M = n_x \times n_y \times N \times D$

*m*=1, 2 ….*D* parameters$c,\beta$

**Output:** motion corrected image $x_0 \in \mathbb{R}^T$ where $T = n_x \times n_y \times N$

---

**Initialization**: Select $X^0 = \begin{bmatrix} x_1 \\ x_2 \\ \vdots \\ x_D \end{bmatrix}$ is a cine image vector at different respiratory states, reconstructed using first stage CS and$\hat{y}^{\mathbf{0}} = y$.

Compute transform coefficients

Index $i$=0

Step-1 (**Demon Registration**: Find $\mathbf{U}^i = \begin{bmatrix} U_1 \\ U_2 \\ \vdots \\ U_D \end{bmatrix}$ using Eq.5

Step-2 (**Solution Update**): $e = \mathbf{U}^{H^i}(\mathrm{F}^H(\hat{y}^i - \mathrm{F}X^i))$ where $X^i = \mathbf{U}^i x_0^i$ and $\mathrm{F}$ is a Fourier operator.

Step-4 Compute sparse coefficients $\mathrm{z}^{\mathrm{i}} = W\left(\dfrac{e}{c} + \mathbf{U}^{\mathrm{H}^{\mathrm{i}}} X^i\right)$

Step-3 **(Soft Thresholding)**:

$$.\hat{\mathbf{z}}_i = \left\{S_{\frac{\beta}{c}}(z^i)\right\} \cong \begin{cases} z^i - \frac{\beta}{c}, & z^i > \frac{\beta}{c} \\ z^i + \frac{\beta}{c}, & z^i < -\frac{\beta}{c} \\ 0, & Otherwise \end{cases}$$

Step-4 $\hat{y}^{\boldsymbol{i}} = \mathrm{F}^H(\mathbf{U}^i(W^{-\mathbf{1}}\hat{\mathbf{z}}_{\mathbf{i}}))$

Step 5 $\hat{y}^{\boldsymbol{i}+\mathbf{1}} = \begin{cases} \hat{y}^{\mathbf{0}} & if\ sample\ is\ acquired \\ \hat{y}^{\boldsymbol{i}}, & otherwise \end{cases}$

Step 6 $X^{i+1} = \mathrm{F}\hat{y}^{\boldsymbol{i}+\mathbf{1}}$

**(Repeat)**: Until stopping criteria

**Stopping Criteria: if** $\left\|\hat{y}^i - \mathrm{F}X^i\right\|_2^2 < tol1$ and $\left\|x_1 - U_m^H x_m\right\|_2^2 < tol2$ stops

**Output** $x_0 = \mathbf{U}^{H^i} X^i$

---

### C. Experimental Data

The proposed method is tested on the simulated free breathing cardiac cine MRI and in-vivo free breathing cardiac cine MRI. A framework for realistic cardiac MR simulations (MRXCAT) was proposed to evaluate the performance analysis of the image reconstruction algorithms from under sampled data in the presence of motion [20]. For simulated data, CINE images with reconstruction matrix size: 256 x256, 24 cardiac phases. The total 2d acquisition was fully sampled cardiac cine images were retrospectively under-sampled by the acceleration rates R= (2, 4, 8, and 12) using $k_y$-t variable-density random under sampling method. Since motion of the heart was primarily in the H-F direction with small non rigid deformation, therefore, small amount of non-rigid deformation noise was added in each cardiac image, by changing orientation of the heart to represent its motion during respiration [21]. The reference respiratory state was the end expiration at which data were acquired.

In free breathing cardiac cine MRI acquisition, ECG-gated acquisition was performed on a Philips 1.5T scanner (b-SSFP), reconstruction matrix size: 160 x160, 24 cardiac phases, with an image resolution of 2.5mm x 2.5 mm x 8mm. The total number of cardiac cycles acquired were 25 cycles, the acquired data were retrospectively under sampled by the acceleration rates R= (2, 4, and 8) using $k_y$-t variable-density random under sampling method. The performance of the proposed framework was compared against the free breathing image reconstructions without motion correction using both simulated and free breathing data.

The separable surrogate functional algorithm solves the minimization problem, illustrated in equation 6. In our simulations, the optimal value of β is found empirically (β=0.0150). The optimal value of c is also found empirically (c= 0.5). MATLAB (The MathWorks, Inc., and Natick, MA) has been used for all implementations.

## III. RESULTS

### A. Simulated Free breathing Data

In the simulations, the performance of the proposed method as a function of the acceleration factor is investigated. Four acceleration rates *R*= (2, 4, 8, and 12) are used to examine the performance of the proposed method. To show the result, three different frames (1, 6, and 12) are picked from the reconstructed cardiac cine MRI.

Figure 3 shows the reconstruction results of the simulated cardiac images at acceleration rates of (2, 4, 8, and 12), without motion correction and the cardiac cine images reconstruction with motion correction. Blurring artifacts are prominent due to non-rigid motion in the CS reconstruction without motion correction for all acceleration rates. At acceleration rate *R* of 2; CS based reconstruction without motion correction has a comparable quality to the motion corrected CS reconstruction, as shown. However, at higher acceleration rates of 4, 8, and 12, CS without motion correction shows poor quality as compared to the motion corrected CS reconstruction, as shown in Figure 3.

### B. Acquired Free breathing Data

In the acquired free breathing data, the performance of the proposed method as a function of different acceleration factors is investigated. Three acceleration rates *R*= (2, 4, and 8) are chosen to investigate the performance of the method. To show the results, three different frames (1, 5, and 12) are chosen from the reconstructed cardiac cine MRI.

Figure 4 shows the reconstruction of the cardiac images at acceleration rates of (2, 4, and 8) without motion correction and the cardiac cine images reconstruction with motion correction. Blurring artifacts are prominent due to non-rigid motion in the CS reconstruction without motion correction for all acceleration rates. At acceleration rate *R* of 2; CS based reconstruction without motion correction has a good quality but still some artifacts are prominent, as shown in Figure 4. However, at higher acceleration rates of 4 and 8, artifacts in CS without motion correction, increase significantly as compared to the motion correction based CS reconstruction.

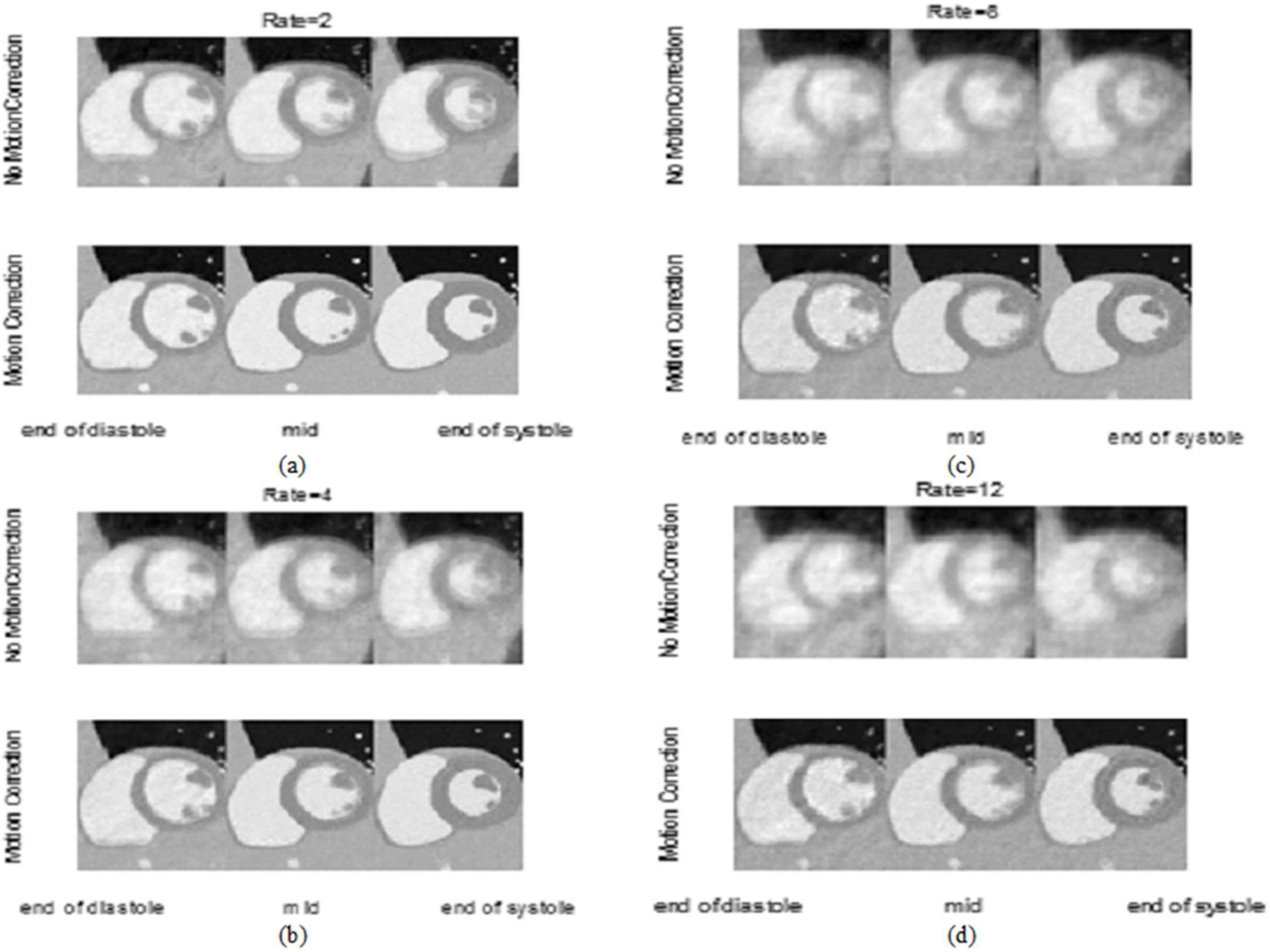

*Figure 3: Performance comparison of our method and CS based reconstruction without motion correction. a) Acceleration rate 2, motion artifact is evident in the upper row as compared to our method. b) Acceleration rate 4, our method still shows better quality c) Acceleration rate 8, without motion correction method deteriorates very rapidly. d) Acceleration 12, under sampling artifact starts appearing in our proposed method.*

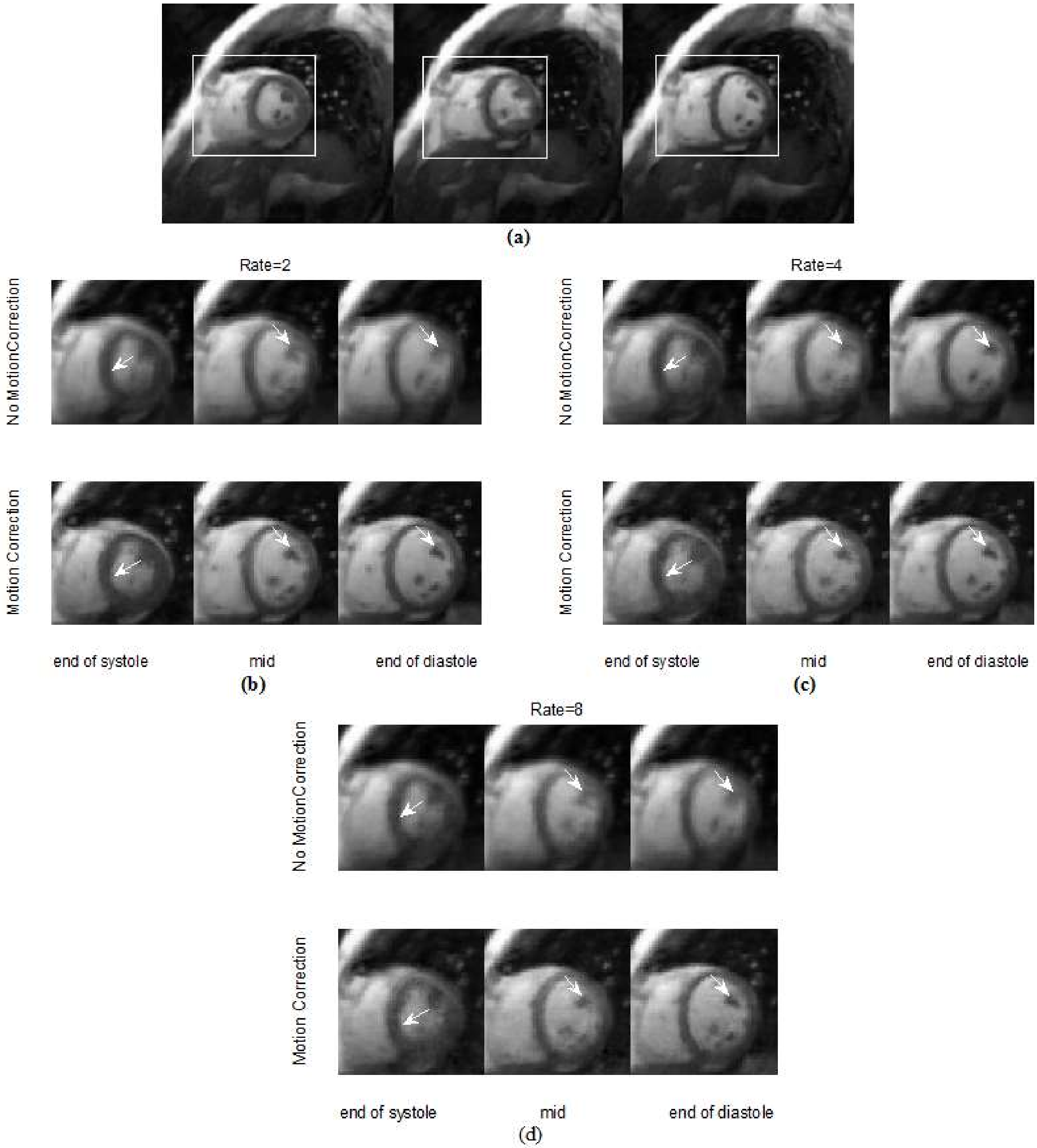

*Figure 4: a) Fully sampled systole, mid and diastole frames with ROI indicated by white box. b) Acceleration rate 2, proposed method provides better quality images, motion artifact is prominent in CS method without motion correction. c) Acceleration rate 4, upper row images deteriorate rapidly as compared to our proposed method. d) Upper row images are of poor quality at high acceleration rate of 8, our proposed method's results are still better as indicated by the bottom row*

.

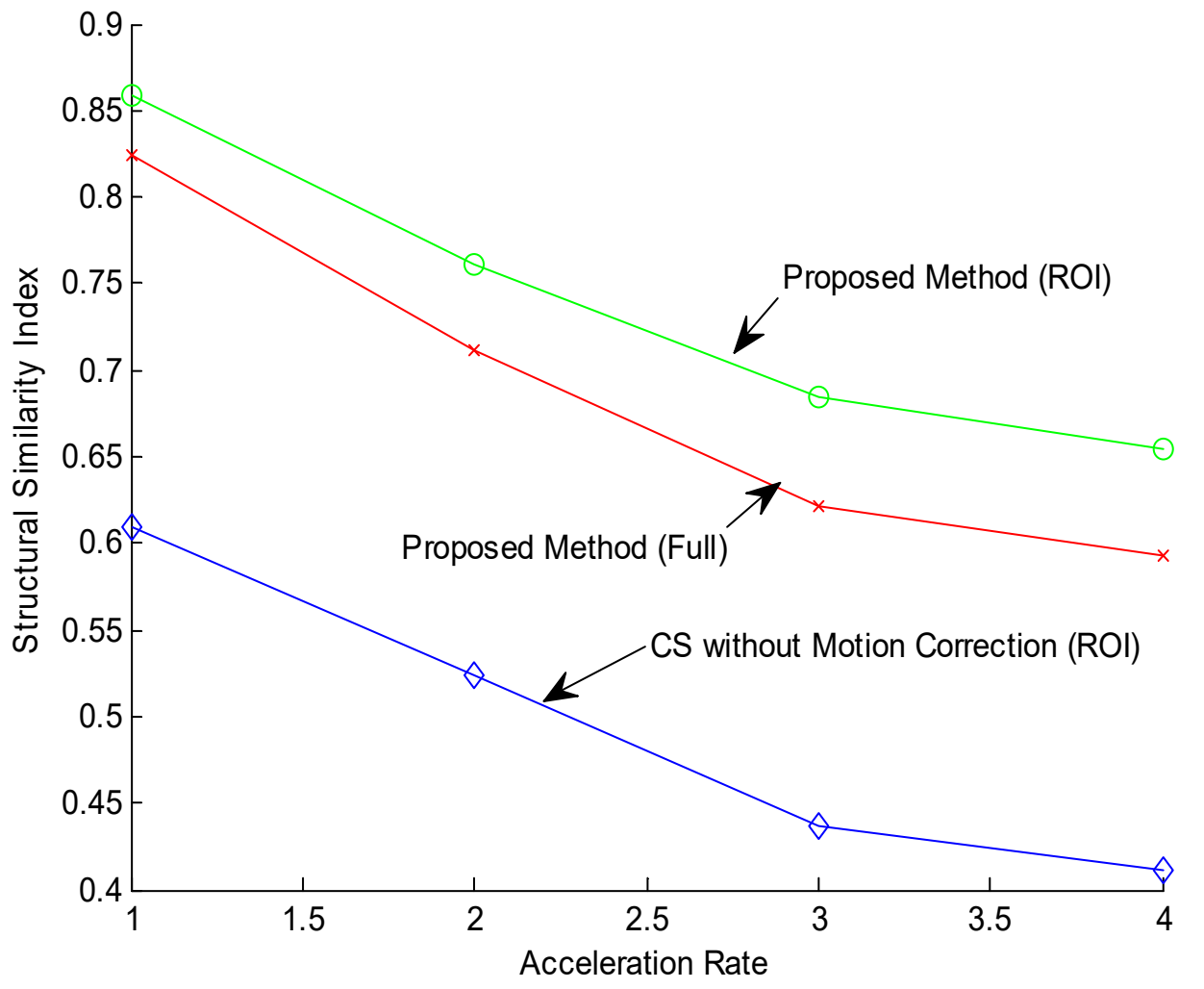

*Figure 5: Performance comparison using structural similarity index. Our method with ROI gives better results (green line), our method with full image (red line) and CS based reconstruction without motion correction (blue line).*

Figure 5 shows the similarity structural index of our motion correction method with ROI, the proposed motion correction method with full image, and the CS based without motion correction method (ROI). This figure depicts better structural index as compared to the CS reconstruction without motion correction. However, our method's structural index deteriorates as the accelerate rate increases. It is because the number of samples to recover the image reduces, and the under sampling artifact becomes prominent.

Figure 6 shows the performance of proposed motion corrected CS technique with different values of beta, at different acceleration rates. Peak signal to noise ratio is used as performance measure, which depicts that the quality of image deteriorates rapidly as the acceleration rate increases. Table 2 shows the performance of proposed motion correction CS reconstruction method as compared to the without motion correction CS reconstruction. The performance parameter used to compare is mean squared error (MSE). Four acceleration rates were used to show the performance of proposed method against the conventional compressed sensing based reconstruction.

*Table 2 Performance Comparison in terms of mean squared error (MSE) using different acceleration rates*

| Acceleration rates *R* | MSE for CS without motion | Proposed method |
|---|---|---|
| 4 | 0.0306 | 0.0207 |
| 8 | 0.0325 | 0.0220 |
| 12 | 0.0337 | 0.0240 |
| 20 | 0.0358 | 0.0270 |

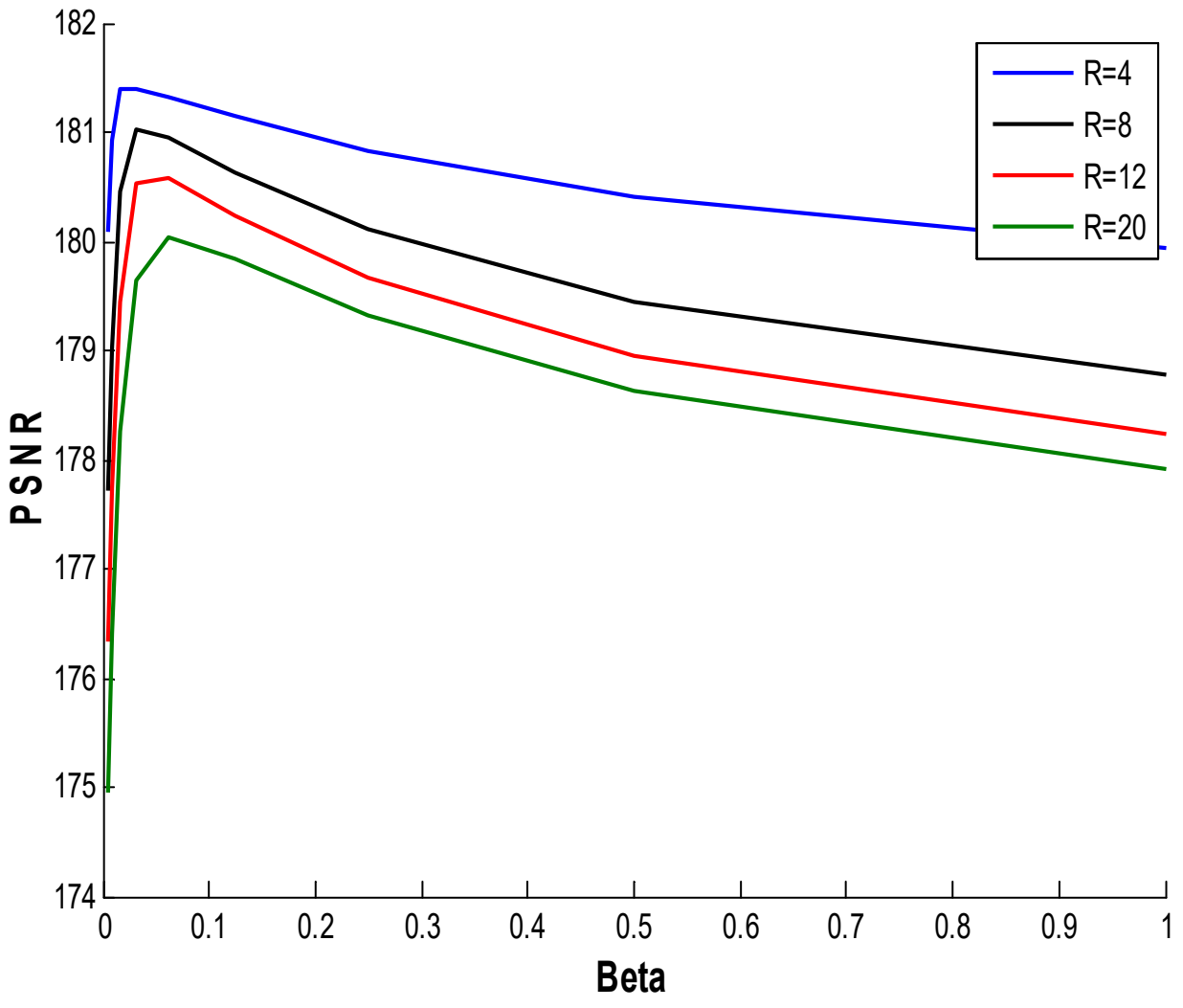

*Figure 6: Performance of proposed reconstruction method using peak signal to noise ratio against different value of beta. The figure shows the range of beta for different acceleration rates at c=0.5*

## IV. Discussion

In this paper, a novel respiratory motion correction based CS reconstruction is proposed. . The data acquired during each cardiac cycle is reconstructed using compressed sensing approach. Images reconstructed in first stage, are sorted using registration based similarity measure, belonging to different respiratory states. Demon based registration is used to estimate the motion between the reference respiratory state and other respiratory states. Finally, separable surrogate functional and motion matrix information are used to reconstruct the motion corrected cardiac cine images. Results show the good quality images as compared to CS reconstructed without motion correction images.

Arbitrary motion correction in dynamic cardiac MRI was presented using radial sampling trajectory [13]. The drawbacks of this approach were addressed in the proposed motion correction method. The changes are mentioned below:

- The variable density Cartesian trajectory is used, which requires uniform Fourier operation as compared to the golden angle radial trajectory which uses non-uniform Fourier operation.
- Binning of data is performed using mutual information based similarity measure without generating the navigator signal from the rigid registration of the low resolution images.
- Separable surrogate functional reconstruction algorithm is used, which fills the missing the samples in the Fourier domain.
- Hierarchical adaptive local affine registration adds artifacts as levels are increased, which is overcome by using the demon based registration algorithms [22].

Recently, cardiac MRI reconstruction was presented to compensate the respiratory artifacts [16]. This technique was computational very complex. Non-uniform Fourier transforms was used, as the radial trajectory was employed for sampling and with increase in respiratory states, the technique takes more computational time. Our proposed scheme was quite simple, with only few processing blocks were involved, as Cartesian variable density sampling was used. As our scheme is less complex with providing good image quality, therefore, high acceleration rates can be employed.

The proposed method gives better results at highly compressive data. However, there are some areas which require further investigations, e.g, different registration based similarity measures need to be studied. More registration algorithms can be investigated, which may perform better in highly under sampled data. Compressed sensing reconstruction algorithm requires further investigations, especially, those improvements which perform well under noisy measurements, exploiting the different sparsities [23]. The proposed method deals the motion artifacts in the normal patients but requires changes to apply to the patients with arrhythmias and other abnormalities.

## V. CONCLUSIONS

A novel motion correction based reconstruction framework is proposed for motion corrupted cardiac cine MRI which achieves motion free images from the highly under-sampled free- breathing data. Simulated and in vivo data show the better performance of our proposed technique as compared to the CS based reconstruction. A proposed compressed sensing based reconstruction is not computational complex, therefore, can be implemented in clinical settings.